\documentstyle[aps,preprint,prb,tighten,floats,psfig]{revtex}
\begin{document}\draft

\title{Reasonable and robust Hamiltonians violating the Third Law}
\author{Greg Watson,$^1$ Geoff Canright$^{2,3}$ and Frank L.\ Somer,
Jr.$^{2,3}$}
\address{\ \\$^{1}$Rutherford Appleton Laboratory\\
Chilton, Didcot, Oxon OX11 0QX, UK\\
$^{2}$Department of Physics and Astronomy\\
The University of Tennessee, Knoxville, TN 37996\\
$^{3}$ Solid State Division, Oak Ridge National Laboratory\\
Oak Ridge, Tennessee 37831}
\maketitle

\begin{abstract}
It has recently been shown that the third law of thermodynamics is
violated by an entire class of classical Hamiltonians in one dimension,
over a finite volume of coupling-constant space, assuming only that
certain elementary symmetries are exact, and that the interactions
are finite-ranged.  However, until now, only the existence of such
Hamiltonians was known, while almost nothing was known of the nature
of the couplings.  Here we show how to define the subvolume of these
Hamiltonians---a `wedge' $W$ in a $d$-dimensional space---in terms of
simple properties of a directed graph.  We then give a simple expression
for a specific Hamiltonian ${\cal H}^*$ in this wedge, and show that
${\cal H}^*$ is a physically reasonable Hamiltonian, in the sense that
its coupling constants lie within an envelope which decreases smoothly,
as a function of the range $l$, to zero at $l=r+1$, where $r$ is the
range of the interaction.
\end{abstract}

\section{Introduction}
It is sometimes stated\cite{barb89} that all materials in their
lowest-energy states are perfect crystals, i.e.~that matter at zero
temperature is characterized by periodic order of the atoms.  If this
is true, it follows that disorder persisting to low temperatures, in
amorphous solids for example, must be interpreted as a result of trapping
of the system in a metastable state.  Another consequence is that the
third law of thermodynamics holds in the Planck form,\cite{aize81,miek87}
which states that the entropy density tends to zero as the temperature
$T\to0$.

These statements, however plausible,\cite{ande84} have not been proved.
One might aim to prove that any physically reasonable microscopic
Hamiltonian describing a material has a unique ground state which is
spatially periodic.  More generally, one wishes to know the minimal
conditions on the Hamiltonian sufficient to guarantee a periodic
ground state.  Both problems are unsolved in general, but some progress
has been made, particularly for one-dimensional systems.\cite{miek87}
Radin and Schulman\cite{radi83} showed that if attention is restricted
to a one-dimensional system of interacting classical units (`spins'),
each of which can exist in a finite number $k$ of distinct states, with
no interactions beyond a spatial range $r$, then there exists a ground
state which is periodic with period at most $k^r$.  In particular,
if the ground state is nondegenerate then it has perfect periodic order.

Thus, in this class of model systems, disorder can occur only if the
ground state is degenerate.  This result can be strengthened by the
observation that degenerate ground states occur only `rarely', in
the sense that they require fine-tuning of the system's parameters
(coupling constants) to precise values.\cite{miek87,teub90,canr96}
In other words, degeneracy occurs only on a set of measure zero in
the space of Hamiltonians.  In the absence of accidental degeneracies,
then, Radin and Schulman's result implies that the ground state of such
a discrete classical system is always periodic.

Recently, however, Canright and Watson\cite{canr96} (CW) have shown
that this picture must be modified if the system is constrained
by an exact symmetry.  The idea that symmetry can imply degeneracy
is familiar.  For the discrete classical chain, CW showed that, under
suitable circumstances, the degeneracy arising from symmetry can result
in a nonzero entropy density, throughout a finite volume of the space of
coupling parameters.  In this phase, termed a D-pair phase, almost all the
ground state configurations are aperiodic.  The D-pair phase is robust, in
the sense that it is not sensitive to small perturbations in the coupling
constants defining the Hamiltonian, as long as these perturbations
respect the symmetry and the restriction to interactions of range $r$.
It is also sufficiently robust to persist to finite temperatures.

CW considered two symmetries in detail: spatial inversion ($I$), and
spin inversion ($S$).  They showed rigorously that, for $S$ symmetry,
D-pair phases exist if and only if $k$ is odd, while for $I$ symmetry,
they exist for $k\ge3$ and $r\ge2$.  The Ising ($k=2$) case is exceptional
in that D-pair phases occur with $I$ symmetry only for range $r\ge5$.

Although the CW proof is constructive, in the sense that it provides
a method for finding all possible D-pair configurations for given $k$
and $r$, there are immediate open questions.  The CW result establishes
existence or nonexistence of D-pairs in each case, but gives no
information on the characteristics of the region in the phase diagram
(the space of coupling parameters) occupied by the D-pair phase, when it
exists---except that it has finite volume.  One would like to know the
size and location of the D-pair region.  The location is important, since
D-pair phases are of little interest unless they occur in a physically
reasonable part of the phase diagram.  For example, consider a Hamiltonian
whose coupling constants increase with spatial separation, and then drop
to zero beyond the cutoff range $r$.  We consider such a Hamiltonian to
be `unphysical'.  Conversely, if the D-pair region includes Hamiltonians
whose couplings decrease smoothly as a function of interaction range $l$,
reaching zero at $l=r+1$ (or before), then we would claim that the case
has been made that physically reasonable Hamiltonians can give ground
states violating the third law.

In this paper we investigate these questions.  After Sec.~\ref{graphs},
which reviews the formalism used in the construction of D-pairs, we derive
in Sec.~\ref{char} results which characterize the geometry of the D-pair
region in terms of the combinatorial properties of the corresponding
graph cycles.  In Sec.~\ref{dph}, we provide a simple construction for
writing down an explicit Hamiltonian corresponding to any given D-pair.
We prove that this Hamiltonian has couplings which fall off approximately
linearly with distance, which shows that it is indeed possible to have
D-pair phases without pathological Hamiltonians.

\section{Graphs, correlation polytopes and D-pairs}\label{graphs}
The system of interest is composed of interacting classical units,
forming an infinite one-dimensional chain.  Each `spin' $\sigma_i$ can
take $k$ distinct values, which we label $0,1,\ldots,k-1$.  A general
Hamiltonian having interactions of maximum range $r$ can be written
\begin{equation}
H = \sum_i f(\sigma_i,\sigma_{i+1},\ldots,\sigma_{i+r}),
\end{equation}
where the sum is over all sites.  Our interest is in ground states of
$H$, which are those configurations $\{\sigma_i\}$ that minimize the
energy density in the thermodynamic limit, i.e., ${\cal H}\equiv H/N$
with the number of sites $N\to\infty$.  In particular, we seek ground
states which do not require fine-tuning of coupling parameters to precise
values; hereafter we restrict the term ground state to mean minimum-energy
states which are robust with respect to small changes in the Hamiltonian.
(A more precise definition in this context is given in
Ref.~\onlinecite{canr96}.)

It is very useful to represent the Hamiltonian pictorially as
a directed graph $G_r^{(k)}$ with energy weights assigned to the
arcs.\cite{miek87,teub90,canr96,bund73} The graph has $k^r$ nodes, each
representing a possible sequence of $r$ spins in the system.  The arcs
in the graph correspond to the operation of spatial translation in the
chain by one unit: a directed arc connects two nodes if the rightmost
$r-1$ spins of one agree with the leftmost $r-1$ spins of the other.
The arc pointing from the node $(\sigma_0,\sigma_1,\ldots,\sigma_{r-1})$
to the node $(\sigma_1,\sigma_2,\ldots,\sigma_r)$ is assigned an energy
weight $f(\sigma_0,\sigma_1,\ldots,\sigma_r)$.  Any spin configuration
of the chain is represented by an infinite path in the graph, and its
energy density equals the average weight of the arcs in the path:
$\epsilon = E/N = \sum_\sigma f(\sigma)n_\sigma$, where $n_\sigma$
is defined as the average occurrence of an arc $\sigma$ in the path.
Thus, each spin configuration is characterized by its arc densities
$\{n_\sigma\}$.  The arc densities are not all independent, since they
satisfy flow constraints\cite{teub90} which state that at each node the
sums of incoming and outgoing arc densities are equal.  In addition,
they satisfy the inequalities $0\le n_\sigma\le1$.

In this language, the Radin--Schulman result is easily understood.
Any path in a graph may be decomposed into simple cycles\cite{robi80}
(SCs), where a SC is a closed path not visiting any node more than once.
If $G_r^{(k)}$ has a unique SC with lower energy per spin than any
other, then the nondegenerate periodic ground state of $H$ is generated
by repetition of that SC; if there are two or more lowest-weight SCs,
then there is always a periodic ground state generated by repeating one
of them.  In either case, the period of the periodic ground state is at
most the number of nodes in $G_r^{(k)}$, which is $k^r$.

That these SCs are true ground states, in our restricted sense of being
stable to perturbations in the Hamiltonian, is readily understood using
the idea of the correlation polytope\cite{teub90,canr96} $P_r^{(k)}$.
The spin correlations are defined by
\begin{equation}
s_\alpha = \langle\sigma_i^{p_0}\sigma_{i+1}^{p_1}\ldots
\sigma_{i+r}^{p_r}\rangle,
\end{equation}
where $\alpha$ denotes the sequence of integers $(p_0,p_1,\ldots,p_r)$;
there are $d=(k-1)k^r$ independent spin correlations, given by the
values $p_i=0,1,\ldots,k-1$ with $p_0\ne0$.  To any configuration of the
chain corresponds a $d$-dimensional vector ${\bf s}$ of correlations,
and any Hamiltonian density can be written as a linear combination,
${\cal H}=-\sum J_\alpha s_\alpha = -{\bf J}\cdot{\bf s}$, where the
$J_\alpha$ are the $d$ independent coupling parameters.  However, the
mapping from configuration to correlation vector is not one-to-one,
and not all correlation vectors represent feasible configurations.
Specifically, the correlations and the arc densities are linearly
related (Sec.~\ref{dph}), and the constraints $0\le n_\sigma\le1$
on arc densities translate to inequalities on the correlation vector.
They constrain ${\bf s}$ to lie inside a convex polytope, and this is
the correlation polytope $P_r^{(k)}$.

Because the Hamiltonian is a linear function of the correlations,
the ground states which are robust to small changes in couplings
$J_\alpha$ are precisely the vertices of $P_r^{(k)}$.  By a simple
argument\cite{teub90,canr96} the vertices can be shown to be in one-to-one
correspondence with the SCs of $G_r^{(k)}$, and we arrive at the result
that the ground states are `almost always' periodic.  One can enumerate
all possible ground states by finding all SCs of the graph.

The argument just sketched does not apply when a symmetry $X$ is
imposed, forcing symmetry-related arc weights to be equal.  If the
lowest-weight SC is not symmetry-invariant, there must be a pair of
degenerate lowest-weight SCs.  If these do not share a node, there exist
two symmetry-broken periodic ground states.  If they share one or more
nodes, the domain wall energy between them is zero, so that there are
infinitely many degenerate ground state configurations, most of which are
mixtures with a nonzero density of domain walls.  The latter case is the
D-pair phase, so called since it comes from a pair of symmetry-broken
configurations, and is characterized by Degeneracy (infinitely many
ground states, yielding a nonzero entropy density) and Disorder (almost
all the ground states have no long-range order).

A simple example\cite{canr96} illustrating the idea of a D-pair is the
$k=3$, $r=1$ model
\begin{equation}
{\cal H} = -\langle\sigma^2\rangle+\langle\sigma_i^2\sigma_{i+1}^2\rangle,
\label{h13}\end{equation}
where the spins $\sigma_i$ can take the values 0 and $\pm1$ and
the angular brackets denote an average over the chain. ${\cal H}$ is
invariant under spin inversion ($S$) symmetry, $\sigma\to-\sigma$.  It is
useful to transform to the variables $\tau=2\sigma^2-1$, which take the
values $\pm1$.  The Hamiltonian becomes, apart from irrelevant constants,
${\cal H}=\langle\tau_i\tau_{i+1}\rangle$, the Ising antiferromagnet.
Its antiferromagnetic ground state, when transformed back to $\sigma$
variables, is $(\ldots{\pm}0{\pm}0\ldots)$, where each $\pm$ spin can
take any value independent of all the others.

The degeneracy and disorder in the ground state appears in this example
as a trivial consequence of the double-valued transformation between
$\tau$ and $\sigma$.  What makes it special is the fact that {\it these
properties are stable to perturbations in the Hamiltonian,} provided these
respect the symmetry and the restriction to range 2 interactions.  This
follows from two facts. (i) The ground states are lower in energy density
than any other (periodic) state by a discrete amount, so a sufficiently
small perturbation cannot create a new ground state. (ii) The only allowed
perturbation terms are those involving correlations already in ${\cal H}$,
plus the additional correlation $\langle\sigma_i\sigma_{i+1}\rangle$, and
each of these takes the same value on all ground states, so the degeneracy
is not split.  This is the robustness characteristic of a D-pair.

\section{Characterizing the D-pair region}\label{char}
To study D-pairs for general $k$ and $r$, CW introduced the concept of
the reduced graph $^X\!G_r^{(k)}$.  Not all symmetry-related pairs of SCs
of $G_r^{(k)}$ correspond to possible ground states in the presence of
symmetry, because the equality of symmetry-related arc weights can imply
the existence of a third SC with lower energy than the original pair.
We refer to this situation as decomposition of a SC pair.  The definitions
of the symmetry-reduced graph $^X\!G_r^{(k)}$ and its SCs are tailored
to take care of decomposing SCs, in such a way that the possible ground
states are in one-to-one correspondence with SCs of $^X\!G_r^{(k)}$.
For $S$ symmetry, the reduced graph is constructed by identifying each
node or arc with its inverse, and SCs are defined as usual as paths
which do not self-intersect.  For $I$ symmetry, the definition of the
reduced graph and its SCs is more involved; we refer the reader to CW
for the technical details, including the classification of SCs into four
topological types.

The reduced graph $^X\!G_r^{(k)}$ allows the enumeration of the ground
state spin configurations for all D-pair phases with a given $k$ and $r$.
Here, we address the question of the region in the phase diagram in
which a given D-pair phase is stable.  By the phase diagram, we mean
the $d^{(X)}$-dimensional space (reduced from $d$ dimensions by the
constraints arising from symmetry) of the coupling parameters $J_\alpha$.

First, let us discuss the problem unconstrained by symmetry.
We ask, what is the region, $W$, of ${\bf J}$-space in which a given
configuration (i.e.~SC) $\omega$ is the ground state?  In principle,
it is a region bounded by hyperplanes corresponding to the inequalities
${\cal H}(\omega)<{\cal H}(\omega')$, where $\omega'$ ranges over all
other SCs.  However, in general some of these inequalities are redundant.
We wish to determine the minimal set of inequalities needed to specify
$W$ fully.  The following two lemmas provide a solution to this problem.

\medskip\noindent
{\bf Lemma 1.} Suppose the SC $\omega$ corresponds to a vertex ${\bf v}$
of the correlation polytope.  The region $W$ of the phase diagram in
which $\omega$ is a ground state is specified by the inequalities ${\bf
J}\cdot({\bf v}-{\bf v}')>0$, where ${\bf v}'$ ranges over the vertices
neighbouring ${\bf v}$, i.e.~those vertices connected in $P_r^{(k)}$
to ${\bf v}$ by a one-dimensional edge.  Furthermore, this set of
inequalities is minimal, in the sense that if any one of them is omitted
the resulting region is strictly larger than $W$.

\medskip\noindent
{\sl Proof:} Let $\{{\bf v}_1,{\bf v}_2,\ldots,{\bf v}_q\}$ be the
vertices of $P_r^{(k)}$ with ${\bf v}_1={\bf v}$, and suppose $\{{\bf
v}_2,{\bf v}_3,\ldots,{\bf v}_p\}$ are the neighbours of ${\bf v}_1$.
Then $W$ is the set of ${\bf J}$ such that ${\bf J}\cdot({\bf v}_i-{\bf
v}_1)<0$ for $2\le i\le q$, and we define $W'$ to be the set of ${\bf
J}$ such that ${\bf J}\cdot({\bf v}_i-{\bf v}_1)<0$ for $2\le i\le p$.
Since $W\subset W'$, to prove $W=W'$ we must show $W'\subset W$.

It follows from the convexity of $P_r^{(k)}$ that the set of vectors
${\bf v}_i-{\bf v}_1$, $2\le i\le p$, from ${\bf v}_1$ to its neighbours
spans the full $d$-dimensional space.  In fact, for $j>p$,
\begin{equation}
{\bf v}_j-{\bf v}_1 = \sum_{i=2}^p\alpha_i({\bf v}_i-{\bf v}_1),
\label{span}\end{equation}
for some $\alpha_i\ge0$, with at least two $\alpha_i\ne 0$.\cite{conv} If
${\bf J}\in W'$, taking its dot product with both sides of (\ref{span})
yields ${\bf J}\cdot({\bf v}_j-{\bf v}_1)<0$, and hence ${\bf J}\in W$,
as required.

Define $W''$ as for $W'$ but omitting one neighbour, say ${\bf v}_2$.
Since ${\bf v}_2$ is a neighbour of ${\bf v}_1$ and since $P_r^{(k)}$
is convex, there exists a hyperplane of dimension $d-1$ intersecting
$P_r^{(k)}$ only in the edge joining ${\bf v}_1$ and ${\bf v}_2$.
Let ${\bf J}$ be a perpendicular vector to this plane from the origin.
The sign of ${\bf J}$ can be chosen so that ${\bf J}\cdot({\bf v}_i-{\bf
v}_1)<0$ for $i>2$, and thus $J\in W''$, while ${\bf J}\cdot({\bf
v}_2-{\bf v}_1)=0$ implies ${\bf J}\not\in W$.
\vrule height5pt width5pt

\medskip\noindent
{\bf Lemma 2.} Two vertices in $P_r^{(k)}$ are neighbours if and only
if the corresponding SCs of $G_r^{(k)}$ have zero or one contacts,
where a contact is a consecutive sequence of one or more shared nodes.

\begin{figure}
\centerline{\psfig
{file=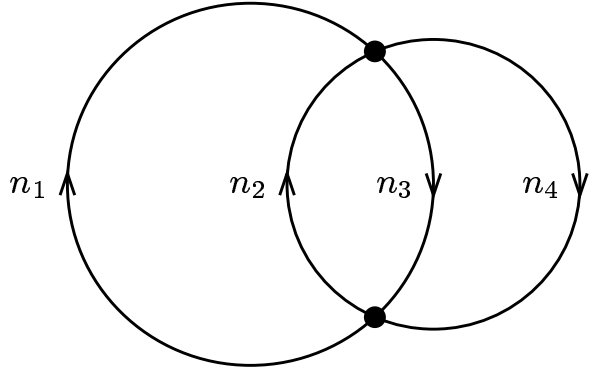,bbllx=150pt,bblly=190pt,bburx=480pt,bbury=410pt,width=8cm}}
\caption{Schematic picture of simple cycles in $G_r^{(k)}$ having two
contacts.  The two circles represent the cycles, made up of arc
sequences labelled with weights $n_1$ to $n_4$, and the large dots are
the contacts, which may consist of more than one node.}\label{decomp}
\end{figure}

\medskip\noindent
{\sl Proof:} The vector of arc densities, ${\bf n}$, has dimension equal
to the number of arcs, but the flow constraints (Sec.~\ref{graphs})
constrain it to lie in a $d$-dimensional subspace which we denote $P'$.
It is the image of $P_r^{(k)}$ under a nonsingular linear transformation
$M$ from ${\bf s}$ to ${\bf n}$ (see Sec.~\ref{dph} for explicit
relations).  It follows that neighbouring vertices of $P_r^{(k)}$
correspond to neighbouring vertices of $P'$.  Two vertices ${\bf v}_1$
and ${\bf v}_2$ are neighbours if and only if any point $\lambda_1{\bf
v}_1+\lambda_2{\bf v}_2$ (with $\lambda_1+\lambda_2=1$) on the line
segment joining them cannot be written as a weighted average of vertices
in any other way.  Suppose two SCs have two contacts, as illustrated
schematically in Fig.~\ref{decomp}.  Recognizing that the four arc
sequences define {\it four} distinct SCs, we can consider a general
convex combination including coefficients $\lambda_3$ for the inner
cycle (arcs 2 and 3) and $\lambda_4$ for the outer cycle (arcs 1 and 4).
A point on the line segment joining the corresponding vertices in $P'$
has densities $n_1=n_3=\lambda$ and $n_2=n_4=1-\lambda$.  Clearly,
there are many convex combinations of vertices yielding the same
densities; for example, if $\lambda<1/2$ we can take $\lambda_1=0$,
$\lambda_2=1-2\lambda$ and $\lambda_3=\lambda_4=\lambda$.  Hence the two
SCs correspond to vertices which are not neighbours.  Conversely, if the
SCs have fewer than two contacts there is only one way to express points
in density space on the segment joining them as a convex combination of
SCs, and so they correspond to neighbouring vertices.
\vrule height5pt width5pt\relax

\medskip
We remark that the reasoning in Lemma 2 is very similar to that leading
to decomposition of pairs of SCs (except that in Lemma 2 the SCs need not
be related by symmetry).  This has a simple geometrical interpretation.
A symmetry imposes linear constraints on the coupling parameters
$J_\alpha$, which means that the relevant space of correlations is a
reduced polytope $^X\!P_r^{(k)}$ obtained by symmetry projection of
$P_r^{(k)}$.  When symmetry-related pairs of vertices of $P_r^{(k)}$
are projected, the ones which become vertices of $^X\!P_r^{(k)}$ are
those which are connected by an edge.  Indeed, the CW definition of SCs
of $^X\!G_r^{(k)}$ (for $X=S$ or $I$) is constructed so as to include
all cycles in $G_r^{(k)}$ which have at most one contact with their
symmetry-partners.  There is, however, one category of non-decomposing
SC pairs which does not correspond to neighbouring vertices.  It has the
form of Fig.~\ref{decomp}, in the case that symmetry forces the weights
to satisfy $w_1=w_2$ and $w_3=w_4$.  This situation occurs with $I$
symmetry for a type four SC, when the two contacts are symmetry-inverses
of each other.  Because of the weight constraints all four cycles in
the diagram have equal energy, and the original pair does not decompose.
It corresponds to two non-neighbouring vertices of a quadrilateral face
of $P_r^{(k)}$, such that all four vertices of the face map to the same
vertex of $^I\!P_r^{(k)}$ under the symmetry.

Lemmas 1 and 2 show how to determine the region in coupling space
corresponding to a given ground state configuration; in fact, they
provide an algorithm for doing this.  From Lemma 1, only neighbouring
vertices need be considered.  Lemma 2 translates the concept of
neighbouring vertices into properties of graph cycles.  In terms of
spin configurations, a contact between SCs means a common string of $r$
or more spins.

Let us now consider the analogous problem in the presence of a symmetry
$X$.  Since Lemma 1 relies only on the Hamiltonian density being a scalar
product, it applies directly to the symmetry-constrained problem, i.e.~the
inequalities defining the stable region $W$ come from neighbouring
vertices of the reduced (projected) correlation polytope $^X\!P_r^{(k)}$.
Lemma 2 also goes through unchanged in the case of $S$ symmetry, since the
definition of SCs for $^S\!G_r^{(k)}$ is the same as that for $G_r^{(k)}$.
However, Lemma 2 does not apply when the symmetry is $I$.

As in the proof of Lemma 2, it is clear that a pair of SCs in
$^I\!G_r^{(k)}$ represent neighbouring vertices in $^I\!P_r^{(k)}$ if and
only if there do not exist two or more new SCs of $^I\!G_r^{(k)}$ using
only arcs from the original pair.  Because SCs for $I$ symmetry may, when
unfolded into $G_r^{(k)}$, represent pairs of intersecting cycles, there
is more freedom to form these new SCs than in the absence of symmetry.
We find that when the intersection does not contain a symmetric node,
then one contact between the original SCs may be enough to imply new SCs.
Specifically, we find the following:

\medskip\noindent
{\bf Lemma 2$'$.} For $I$ symmetry, two SCs of $^I\!G_r^{(k)}$ correspond
to neighbouring vertices if and only if one of the following conditions
is satisfied: (i) they have no contacts; (ii) they have one contact and
one of them is of type one (i.e.~unfolds to nonintersecting cycles);
(iii) they have one contact which includes a symmetric node.  (In the
last case, both SCs must be type two or three.)

\medskip
Let us illustrate our results with the example of $k=3$ and $r=2$ for
both $S$ and $I$ symmetries.  As in Sec.~\ref{graphs} we shall take the
allowed spin values to be $\sigma=0$ and $\pm1$.

\begin{figure}
\centerline{\psfig
{file=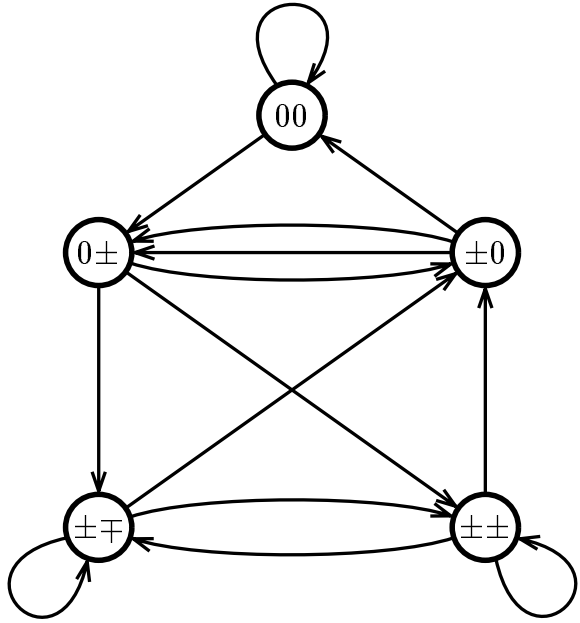,bbllx=80pt,bblly=90pt,bburx=390pt,bbury=420pt,width=8cm}}
\caption{The reduced graph $^S\!G_2^{(3)}$.}\label{sg23}
\end{figure}

Fig.~\ref{sg23} shows the graph $^S\!G_2^{(3)}$.  It has 14 arcs, and 5
nodes each of which implies a flow constraint, leaving 9 independent arc
densities, i.e.~$d^{(S)}=9$.  The 9 symmetry-invariant correlations are
$s_1=\langle\sigma^2\rangle$, $s_2=\langle\sigma_i\sigma_{i+1}\rangle$,
$s_3=\langle\sigma_i\sigma_{i+2}\rangle$, $s_4=\langle\sigma_i^2
\sigma_{i+1}^2\rangle$, $s_5=\langle\sigma_i^2\sigma_{i+2}^2\rangle$,
plus four correlations involving three spins; the Hamiltonian density
is written in terms of its 9 coupling parameters as ${\cal H}=-\sum
J_\alpha s_\alpha $.  The graph has 19 distinct SCs, 5 of which are
D-pairs using the invariant node $(00)$.  For example, let us consider
the D-pair SC $\omega=(00\pm)$.  To find its stable region, we need only
consider the 10 SCs which represent neighbours of $\omega$, according
to Lemma 2.  One neighbouring SC is the ferromagnetic state $(00)$;
comparing its energy to that of $\omega$ yields the condition $J_1>0$,
so we may set $J_1=1$.  For each of the 9 other neighbouring SCs one
can write down the corresponding inequality on the 8 remaining couplings
directly from the graph.  We do not list them here; let us merely display
a typical solution:
\begin{equation}
{\cal H} = -\langle\sigma^2\rangle+\langle\sigma_i^2\sigma_{i+1}^2\rangle
+\langle\sigma_i^2\sigma_{i+2}^2\rangle.
\end{equation}
As in the example of Sec.~\ref{graphs} it is informative to perform the
transformation $\tau=2\sigma^2-1$ to Ising spins.  The Hamiltonian becomes
\begin{equation}
{\cal H} = \langle\tau_i\tau_{i+1}\rangle+\langle\tau_i\tau_{i+2}\rangle
+2\langle\tau\rangle,
\end{equation}
which represents an Ising model with antiferromagnetic nearest and
next-nearest neighbour interactions and a magnetic field favouring the
$-$ state.  It is easy to check that the ground state is $(-{-}+)$,
or in $\sigma$ variables, $(00\pm)$, as required.  Thus we have the
degeneracy and disorder characteristic of a D-pair.  Its robustness
to perturbations in the Hamiltonian follows from the fact that all 9
correlations take identical values on every degenerate configuration;
i.e.~no perturbation made up of $s_1$ to $s_9$ can split the degeneracy.

\begin{figure}
\centerline{\psfig
{file=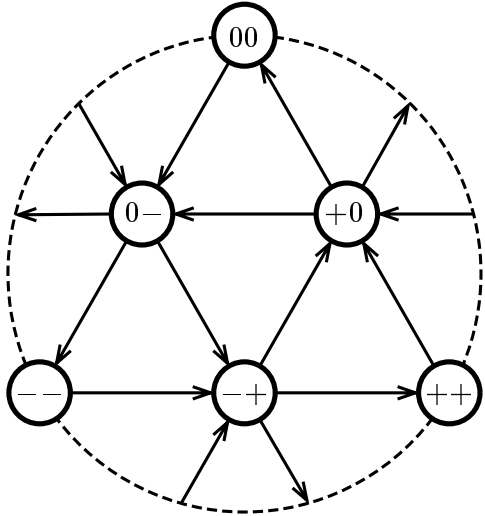,bbllx=135pt,bblly=210pt,bburx=410pt,bbury=500pt,height=7cm}}
\caption{The reduced graph $^I\!G_2^{(3)}$.}\label{ig23}
\end{figure}

The corresponding reduced graph for $I$ symmetry is shown in
Fig.~\ref{ig23}.  For ease in drawing we have distorted the symmetry line
${\cal I}$ into a circle and omitted the ferromagnetic arcs joining each
$I$-invariant node to ${\cal I}$.  This graph has 32 distinct SCs, most
of which are of type three, unfolding to invariant cycles in $G_2^{(3)}$.
There are three (type two) D-pairs, of which we shall consider the example
$(00{-}+)/(00{+}-)$.  It has 15 neighbours according to Lemma $2'$,
namely the ferromagnetic SCs, the type one SC $(0{+}-)/(0{-}+)$, and the
10 type three SCs which use the symmetric node $(00)$.  Thus there are
15 inequalities constraining the 14 distinct $I$-invariant correlations.
Again, we do not list them here, but simply display a particular solution,
which happens to involve only $S$-invariant pairwise interactions:
\begin{equation}
{\cal H} = \langle\sigma_i\sigma_{i+1}\rangle
+\langle\sigma_i^2\sigma_{i+2}^2\rangle.
\end{equation}
The ground state is $(\ldots00{\pm}{\mp}00{\pm}{\mp}00\ldots)$, where
the spins in each $(\pm\mp)$ segment may be chosen independently to be
$({+}-)$ or $({-}+)$.  Once again, one can check that this degeneracy
is not split by any of the 14 possible $I$-invariant perturbing terms
that may be added to the Hamiltonian.

\section{Explicit D-pair Hamiltonian}\label{dph}
The techniques described in the previous section can be used to find the
region of stability corresponding to any given D-pair, for $S$ or $I$
symmetry.  However, the analysis becomes tedious for large $k$ and $r$.
In this section, we describe a simple construction for finding a single
point, ${\cal H}^*$, in the stable region.

The constuction is based on the observation that the arc weights, which
determine the Hamiltonian, can all be chosen independently, i.e.~for
any choice of arc weights there exists a corresponding Hamiltonian.
(This statement should not be confused with the fact that the arc
{\it densities} are not independent because of the flow constraints,
and hence that the map from arc weights to couplings $J_\alpha$ is not
one-to-one.)  Given a D-pair defined by a SC $\omega$ of $^X\!G_r^{(k)}$,
we define ${\cal H}^*$ as the Hamiltonian corresponding to the following
assignment of arc weights: $w_\tau$ is $-1$ if the arc $\tau$ occurs
in the unfolding of the D-pair SC into two intersecting cycles in
$G_r^{(k)}$, and 0 otherwise.  In terms of arc densities, ${\cal H}^*$ is
\begin{equation}
{\cal H}^* = -\sum_{\tau\in\omega}(n_\tau+n_{\bar\tau}),
\label{hstar}\end{equation}
where the overbar denotes symmetry inversion.  By construction, the given
D-pair phase is the ground state of ${\cal H}^*$: since the energy is the
average arc weight, any of the D-pair spin configurations has energy $-1$,
while other configurations use some weight 0 arcs and have higher energy.

In (\ref{hstar}), ${\cal H}^*$ is written in terms of arc
densities.  The latter are related to the correlations as follows.
If $\tau=(\tau_0,\tau_1,\ldots,\tau_r)$, then
\begin{equation}
n_\tau = \langle\delta_{\sigma_i\tau_0}\delta_{\sigma_{i+1}\tau_1}\ldots
\delta_{\sigma_{i+r}\tau_r}\rangle.
\label{dens}\end{equation}
The Kronecker delta, as a function of a spin variable $\sigma$, can be
written as a degree $k-1$ polynomial according to
\begin{equation}
\delta_{\sigma\eta} = \prod_{\eta'\ne\eta}(\sigma-\eta')/(\eta-\eta')
=\sum_{p=0}^{k-1}\chi_{p\eta}\sigma^p.\label{delta}
\end{equation}
The product is over all spin values $\eta'$ not equal to $\eta$, and the
second equality defines the numbers $\chi_{p\eta}$ as the coefficients in
the polynomial expansion of the product.  When this is substituted into
(\ref{dens}) the arc densities are given as a linear combination of spin
correlations.  From (\ref{hstar}), this yields an expression for ${\cal
H}^*$ in terms of the correlations.  As an example, when this procedure
is applied to the D-pair $(\pm0)$ for $k=3$, $r=1$ with $S$ symmetry,
the result is the Hamiltonian (\ref{h13}), discussed in Sec.~\ref{graphs}.

Let us investigate further the structure of ${\cal H}^*$.  Its expression
in terms of correlations, from substituting (\ref{delta}) and (\ref{dens})
into (\ref{hstar}), is
\begin{equation}
{\cal H}^* = -\sum_{j=1}^p\sum_{p_0=0}^{k-1}\sum_{p_1=0}^{k-1}\ldots
\sum_{p_r=0}^{k-1} \chi_{p_0\tau_j}\chi_{p_1\tau_{j+1}}\ldots
\chi_{p_r\tau_{j+r}} \langle\sigma_i^{p_0}\sigma_{i+1}^{p_1}\ldots
\sigma_{i+r}^{p_r}\rangle+\hbox{s.i.}
\label{hexp}\end{equation}
Here, $\tau_j$ denotes the $j$th spin (using any arbitrary
starting point) of the configuration defined by the SC $\omega$,
and $p$ is the period of $\omega$.  The second term, not
written explicitly, is the symmetry inverse of the first---that
is, every $\tau\mapsto\overline{\tau}$.  Consider a range $l$
correlation $s_\alpha=\langle\sigma_i^{q_0}\sigma_{i+1}^{q_1}\ldots
\sigma_{i+l}^{q_l}\rangle$, where $q_0$ and $q_l$ are both nonzero.
If $l<r$, there are $(r-l+1)$ terms in (\ref{hexp}) contributing to
$J_\alpha$, the coupling parameter multiplying $s_\alpha$ in ${\cal H}^*$.
We find
\begin{equation}
J_\alpha = \sum_{j=1}^p\sum_{m=0}^{r-l}\chi_{0\tau_j}\ldots
\chi_{0\tau_{j+m-1}}\chi_{q_0\tau_{j+m}}\ldots\chi_{q_l\tau_{j+m+l}}
\chi_{0\tau_{j+m+l+1}}\ldots\chi_{0\tau_{j+r}}+\hbox{s.i.}
\label{jsum}
\end{equation}
The structure of this expression is seen most clearly if we consider
initially the case of Ising spins, $k=2$.  Taking the allowed spin values
to be $\sigma=\pm1$, we have $\chi_{0\eta}=1/2$ and $\chi_{1\eta}=\eta/2$,
and we arrive at the result
\begin{equation}
J_\alpha = 2^{-(r+1)}(r-l+1)\sum_{j=1}^p\tau_j^{q_0}\tau_{j+1}^{q_1}\ldots
\tau_{j+l}^{q_l}+\hbox{s.i.}
\end{equation}
This result is quite significant.  It says that the value of $J_\alpha$
is, apart from a constant $2^{-(r+1)}$ that we shall ignore, equal to
$(r-l+1)[t_\alpha + \rm{s.i.}]$, where $t_\alpha$ is the correlation
$s_\alpha$ evaluated in the spin configuration $\{\tau_i\}$ of the D-pair.

Since a correlation for Ising spins has magnitude at most 1, this implies
the bound $|J_\alpha|\le(r-l+1)$.  Further, we note that $t_\alpha$ is
expected not to be strongly dependent on the range $l$ of the coupling
$J_\alpha$.  Roughly speaking, spin configurations that have long-range
correlations tend also to be correlated at short range.  This idea is
borne out by explicit computations; for instance, for the $I$-symmetry
$r=5$ Ising D-pair $(+{-}{-}{-}{+}{+}-)$, there are 23 symmetric
correlations, each of which takes one of the values $-1/7$ or $3/7$,
with no systematic dependence on $l$.  Thus, the dominant contribution to
$J_\alpha$ comes from the factor $(r-l+1)$.  As a function of distance,
this represents a linear decrease to zero at the cutoff range $l=r+1$.

For $k>2$, the situation is similar.  Of course, the values of the
couplings depend on the choice of the set of allowed spin values, which
has been left arbitrary so far.  However, for any $k$ there exists a
choice with the property that $\chi_{0\eta}$ is independent of $\eta$,
namely, letting $\{\sigma\}$ be the complex $k$th roots of unity.
For this choice, the sum (\ref{jsum}) simplifies as it did for Ising
spins, yielding (apart from a constant) $J_\alpha=(r-l+1)[t_\alpha +
{\rm s.i.}]$.  Under the assumption that the correlations $t_\alpha$
depend weakly on $l$, we find again that the dominant distance dependence
of $J_\alpha$ is a linear fall-off to zero beyond the cutoff range.

\section{Discussion}
The existence of D-pair phases is interesting from a theoretical point
of view.  However, we are not aware of any obvious candidate material for
their realization in nature.  We note that there is an entire class of
materials, namely, layered solids or polytypes, which are well modelled
by effective Hamiltonians such as those studied here.  This class of
materials is however quite large; and the few effective Hamiltonians
that are known from this class do not show promise of having a D-pair
phase as the ground state.  (See the discussion and references in
Ref.~\onlinecite{yi96}.)

Thus, there are significant obstacles to finding D-pairs in practice.
However, the results of Sec.~\ref{dph} of this paper remove one potential
obstacle: the possibility that the only Hamiltonians exhibiting D-pair
phases are pathological in the dependence of their coupling parameters
on distance.  We would like the couplings to decrease smoothly to zero at
the cutoff range, otherwise it would seem unphysical to impose a rigid
cutoff beyond which there are no interactions.  We have constructed
an explicit Hamiltonian for arbitrary $k$ and $r$, which has D-pair
ground states.  Encouragingly, its couplings are very well behaved:
as a function of distance, they fall linearly to zero at the cutoff.

Remaining obstacles concern the robustness of D-pair behaviour.
Although D-pairs are not destroyed by symmetric perturbations of
sufficiently short range or by nonzero temperature, they are in general
destroyed by including interactions beyond the cutoff.  They are also
destroyed by deviations from perfect symmetry caused, for example, by
external fields---which may or may not be strictly zero, depending on
the symmetry in question, and on the physical identity of the `spins'.
However, even when the degeneracy is broken in such ways, behaviour
characteristic of D-pairs may be observable at a suitable energy scale.
If the perturbations breaking the D-pair symmetry are small, they are
not manifest except at very small temperatures.  At low but nonzero
temperatures, one would still expect to observe disordered states and
nonzero entropy density.  In that case, the techniques of Sec.~\ref{char}
of this paper apply directly to the problem of characterizing the
D-pair region.  (Possible experimental signatures of D-pairs have been
investigated by Yi and Canright.\cite{yi96})

Another potential obstacle is the limitation to problems involving
classical, discrete units.  However, such models are likely to be good
approximations for certain problems, such as stacking polytypes of
crystals (see CW and references therein) where the `spin' represents
the discrete set of possible configurations of a single lattice plane.
Another limitation is the restriction to one-dimensional models.  The
question of whether similar behaviour is possible in higher dimensions is
unexplored, although certain frustrated two-dimensional models are known
to have degenerate ground states of large periodicity.\cite{teub96}
Finally, although we have shown that D-pair phases are possible
with Hamiltonians that are not obviously unphysical, there may be
more subtle physical reasons---arising, say, from quantum-mechanical
considerations---which may argue against effective classical Hamiltonians
having D-pairs as ground states.  For example, effective classical
Hamiltonians representing the binding energy of mobile electrons in an
ionic background tend to favour periodic ionic arrangements.\cite{wats95}
We leave these questions for future work.

\bigskip
{\it Acknowledgements.} We thank Ken Stephenson for helpful discussions of
high-dimensional geometry.  This work was supported in part by the U.S.
Department of Energy through Contract No.~DE-AC05-84OR21400 with Martin
Marietta Energy Systems Inc.  GSC and FLS acknowledge support from the
NSF under Grant No.~DMR-9413057.

\end{document}